\documentclass[prb,twocolumn,tbtags,superscriptaddress,floatfix]{revtex4}
\usepackage{amssymb}
\usepackage{graphicx,amsmath}
\usepackage{bm}
\usepackage{times}

\begin{document}

\title{Delayed-response quantum back-action in nanoelectromechanical systems}

\begin{abstract}
We present a semiclassical theory for the delayed response of a quantum dot
(QD) to oscillations of a coupled nanomechanical resonator (NR). We prove
that the back-action of the QD changes both the resonant frequency and the
quality factor of the NR. An increase or decrease in the quality factor of
the NR corresponds to either an enhancement or damping of the oscillations,
which can also be interpreted as Sisyphus amplification or cooling of the NR
by the QD.
\end{abstract}

\date{\today }
\author{S. N. Shevchenko}
\affiliation{B. Verkin Institute for Low Temperature Physics and Engineering, Kharkov,
Ukraine}
\affiliation{V. Karazin Kharkov National University, Kharkov, Ukraine}
\affiliation{CEMS, RIKEN, Saitama, 351-0198, Japan}
\author{D. G. Rubanov}
\affiliation{V. Karazin Kharkov National University, Kharkov, Ukraine}
\author{Franco Nori}
\affiliation{CEMS, RIKEN, Saitama, 351-0198, Japan}
\affiliation{Physics Department, University of Michigan, Ann Arbor, MI 48109-1040, USA}
\maketitle


\section{Introduction}

An important model hybrid system is a resonator coupled to a mesoscopic
normal or superconducting system [\onlinecite{Xiang13}]. In many cases, the
resonator, which can be electrical or nanomechanical, is slow and can be
described classically. This implies the relation $\hbar \omega _{0}<k_{%
\mathrm{B}}T$ between its resonant frequency $\omega _{0}$ and the
temperature $T$. In contrast to this, the characteristic energy of a
mesoscopic quantum subsystem is usually larger than $k_{\mathrm{B}}T$. In
this case, the resonator and the quantum subsystem evolve on different
timescales. Adjustment should be made if, in addition, there is a slow
component in the evolution of the quantum system. One such situation takes
place~[\onlinecite{Hauss08, Greenberg09}] if the Rabi oscillations are
induced with a frequency $\Omega _{R}\sim \omega _{0}$, resulting in an
effective energy exchange between the subsystems.

Another interesting situation occurs when the relaxation of the quantum
subsystem is so slow that its characteristic time $T_{1}$ is of the order of
the resonator's period $T_{0}=2\pi /\omega _{0}$, which is a realistic
assumption for quantum dots~[\onlinecite{Wang13}]. Then the delayed response
of the quantum subsystem to the resonator's perturbation implies that the
resonator is influenced by both the in-phase and out-of-phase forces~[%
\onlinecite{Braginsky74, Metzger04, Clerk05, Bode12}]. The out-of-phase
force can damp or amplify the resonator oscillations [\onlinecite{Xue07}].
Such effects can be described as a decrease or increase in the number of
photons in the resonator, which relates to lasing and cooling~[%
\onlinecite{Armour04, Naik06, Schliesser06, Brown07, Ouyang09,
Ashhab09}].

Alternatively, the slow evolution of a quantum subsystem subject to a
periodic driving by a resonator with a significant probability of relaxation
can be described in terms of periodic \emph{Sisyphus-type processes}. This
was studied for an electric resonator coupled to a superconducting qubit~[%
\onlinecite{Grajcar08, Nori08, Persson10, Skinner10}]. In such systems, the
electric resonator performs Sisyphus-type work by slowly driving a qubit
along a continuously ascending (or descending) trajectory in energy space,
while the cyclic Sisyphus destiny is completed by resonant excitation on one
side of the trajectory and relaxation on the other~[\onlinecite{Nori08}].
Our aim in this paper is to study an analogous process for a typical
nanoelectromechanical system~[\onlinecite{Poot12, Greenberg12}], which
consists of a nanomechanical resonator (NR) coupled to a single-electron
transistor or a quantum dot (QD)~[%
\onlinecite{Irish03, Blencowe05, LaHaye09,
Shekhter13, Benyamini14}]. This study is partly motivated by the experiments
in Refs.~[\onlinecite{Grajcar08,
OkazakiISNTT13}].

A straightforward approach for describing a slow classical resonator coupled
to a fast quantum subsystem is a fully-quantum description of the coalesced
system~[\onlinecite{Hauss08, Grajcar08}]. Arguably, a more intuitively clear
procedure assumes a delayed response of the quantum subsystem to the
resonator driving. The effectiveness of this\textit{\ }\emph{%
delayed-response method} has been confirmed in different contexts~[%
\onlinecite{Braginsky74, Metzger04, Clerk05, Xue07, ShevchenkoPG08,
DeLiberato10}]. In particular, the observation of Sisyphus cooling and
amplification of an electrical \textit{LC} circuit by a flux qubit~[%
\onlinecite{Grajcar08}] can be described by solving the master equation of
the coalesced system [\onlinecite{Hauss08, Grajcar08}]; the delayed-response
method performs equally well in describing such a system~[%
\onlinecite{ShevchenkoPG08,
ShevchenkoIO12}]. In both cases, successful fitting of the experimental
results yields a similar value for the key delay parameter, $\omega
_{0}T_{1}\approx 1$, close to the optimal value for Sisyphus cooling and
amplification.

Accordingly, for a coupled slow classical resonator and a fast quantum
subsystem, we will use a semiclassical theory within the framework of the
delayed-response method. The resonator (here a NR) slowly drives the quantum
subsystem (a QD, in our case), with the response of the latter at a time $t$
determined by the driving parameters at some prior time $\widetilde{t}%
=t-\tau $. We will show that this produces an out-of-phase force, with the
resonator's oscillations amplified or attenuated by the back-action of this
force. While we leave the detailed discussions for the Appendixes, in the
rest of the paper we consider in detail the delayed response of the QD to
the oscillations of the coupled NR. The presentation is organized in such a
way that the approach could be straightforwardly adapted to other similar
systems, where a slowly-driven system is coupled to a fast quantum system,
whose back-action is delayed by the (possibly slow) relaxation process.

\section{Semiclassical theory for the coupled quantum dot and nanomechanical
resonator system}

\subsection{Model}

\begin{figure}[t]
\includegraphics[width=8 cm]{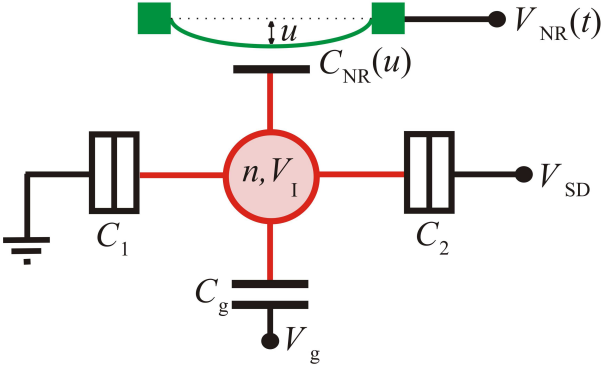}
\caption{(Color online) Schematic diagram of a system composed of a
nanomechanical resonator (green) electrostatically-coupled to a quantum dot
(red). The source (left) and drain (right) electrodes of the QD are biased
by the voltage $V_{\mathrm{SD}}$; the QD state is controlled by the gate
voltage $V_{\mathrm{g}}$. The NR is actuated by the voltage $V_{\mathrm{NR}%
}(t)=V_{\mathrm{NR}}+V_{\mathrm{A}}\sin \protect\omega _{0}t$. The coupling
between the NR and the QD is characterized by the displacement-dependent
capacitance $C_{\mathrm{NR}}(u)$. }
\label{Fig:scheme}
\end{figure}

A schematic diagram for a coupled QD-NR system, analogous to a feasible
experimental setup~[\onlinecite{OkazakiISNTT13, Okazaki13}], is shown in
Fig.~\ref{Fig:scheme}. Here, the essential element is the island or quantum
dot (QD). It is characterized by the total capacitance $C_{\Sigma
}=C_{1}+C_{2}+C_{\mathrm{g}}+C_{\mathrm{NR}}$, average number of excessive
electrons $\left\langle n\right\rangle $, and the island's potential $V_{%
\mathrm{I}}$. The QD is biased by the gate voltage $V_{\mathrm{g}}$ and the
voltage $V_{\mathrm{NR}}$ applied via the capacitance $C_{\mathrm{NR}}(u)$,
one of the plates of which is able to perform mechanical oscillations. This
is the NR, and its displacement $u$ is related to the current through the QD.

Consider the mechanical resonator as a beam with mass $m$, elasticity $k_{0}$%
, and damping factor $\lambda _{0}$ (which is assumed to be small). The
oscillator has an eigenfrequency $\omega _{0}=\sqrt{k_{0}/m}$ and quality
factor $Q_{0}=m\omega _{0}/\lambda _{0}$. The oscillator is assumed to be
driven by the probe periodic force $F_{\mathrm{p}}\sin \omega _{0}t$ but its
state is also influenced by the quantum subsystem, QD, through the force $F_{%
\mathrm{q}}$. This external nonlinear force $F_{\mathrm{q}}$ is taken to
depend only on the position variable $u$ and its derivative, $F_{\mathrm{q}%
}=F_{\mathrm{q}}(u,\dot{u})$. Accordingly, the displacement $u$ is the
solution of the equation of motion~[\onlinecite{Poot12}]
\begin{equation}
m\overset{\cdot \cdot }{u}+\frac{m\omega _{0}}{Q_{0}}\dot{u}+m\omega
_{0}^{2}u=F_{\mathrm{q}}\left( u,\dot{u}\right) +F_{\mathrm{p}}\sin \omega
_{0}t.  \label{(1)}
\end{equation}%
In general, for small oscillations%
\begin{equation}
F_{\mathrm{q}}\left( u,\dot{u}\right) \approx F_{\mathrm{q}0}+\frac{\partial
F_{\mathrm{q}}}{\partial u}u+\frac{\partial F_{\mathrm{q}}}{\partial \dot{u}}%
\dot{u}\;.  \label{small_oscs}
\end{equation}%
It follows that the second term above shifts the elasticity coefficient $%
k_{0}=m\omega _{0}^{2}$ and the resonant frequency $\omega _{0}$ to the
effective frequency $\omega _{\mathrm{eff}}$,%
\begin{equation}
\omega _{\mathrm{eff}}^{2}=\omega _{0}^{2}-\frac{1}{m}\frac{\partial F_{%
\mathrm{q}}}{\partial u}\,,
\end{equation}%
while the third term changes the damping factor $\lambda _{0}=m\omega
_{0}/Q_{0}$, producing an effective quality factor $Q_{\mathrm{eff}}$\
satisfying%
\begin{equation}
\frac{1}{Q_{\mathrm{eff}}}=\frac{1}{Q_{0}}-\frac{1}{m\omega _{0}}\frac{%
\partial F_{\mathrm{q}}}{\partial \dot{u}}\;.  \label{Q}
\end{equation}%
From these results, the expressions for the small frequency shift ($\Delta
\omega \ll \omega _{0}$) and the quality factor shift ($\Delta Q\ll Q_{0}$)
become:%
\begin{equation}
\Delta \omega \equiv \omega _{\mathrm{eff}}-\omega _{0}\approx -\frac{1}{%
2m\omega _{0}}\frac{\partial F_{\mathrm{q}}}{\partial u}\;,  \label{Dw}
\end{equation}%
\begin{equation}
\Delta Q\equiv Q_{\mathrm{eff}}-Q_{0}\approx \frac{Q_{0}^{2}}{m\omega _{0}}%
\frac{\partial F_{\mathrm{q}}}{\partial \dot{u}}\;.  \label{DQ}
\end{equation}

There are various possible scenarios under which this back-action shift of
the qualify factor $\Delta Q$ becomes non-trivial. For example, the
dependence $F_{\mathrm{q}}=F_{\mathrm{q}}(\dot{u})$ could originate from
external forces, as is the case in Ref.~[\onlinecite{Poot10}].
Alternatively, non-trivial $\Delta Q$ also results when there is a lag in
the back-action. Here we consider this latter case in detail.

\subsection{Lagged back-action}

If all the characteristic times of the QD are much faster than those of the
NR, then its back-action is characterized by $F_{\mathrm{q}}=F_{\mathrm{q}%
}\left( u\right) $ and no changes in $Q$ are expected. However, in the next
approximation, the QD sees the dependence $u=u(t)$ and we have $F_{\mathrm{q}%
}=F_{\mathrm{q}}\left( u,\dot{u}\right) $. An illustrative way to describe
this is by phenomenologically introducing a delayed time-dependence in the
QD response to the influence of the NR. This key assumption is discussed in
detail in Appendix A. The delayed-response method can be formulated as
follows.

We assume that without backaction the force is linear in the NR
displacement,
\begin{equation}
F_{\mathrm{q}}=F_{\mathrm{q}0}+\Xi \,u.  \label{F_with_Ksi}
\end{equation}%
Then the delayed time-dependence is characterized by replacing $t\rightarrow
\widetilde{t}=t-\tau $. Here $\tau $ stands for the characteristic time,
which in our case describes the delay needed for changes in $C_{\mathrm{NR}%
}(u)$ to affect the current $I$ in the QD. There are two possible origins of
the delayed response. The first relates to the tunneling rate $\Gamma $,
with a delay time between the in- and out- tunneling events known as the
Wigner-Smith time, $\tau \sim 1/\Gamma $ [%
\onlinecite{Ringel08, Gardner11,
Yin14}]. The second origin of the delayed response is when the upper-level
occupation is created by any means, and the relaxation from it to the ground
state has a delay $\tau \sim T_{1}$ [\onlinecite{Wang13}]. This latter case
is considered in detail in Appendix A.

The delayed-response assumption means that the back-action of the QD is
described by the displacement which defined the position of the NR some time
ago: $F_{\mathrm{q}}(t)=F_{\mathrm{q}}\left[ u(t-\tau )\right] $. For the
induced NR oscillations, $u(t)=v\cos (\omega _{0}t+\delta )$, we then have
\begin{equation}
u(t-\tau )=v\,\mathcal{C}\cos (\omega _{0}t+\delta )+v\,\mathcal{S}\sin
(\omega _{0}t+\delta )  \label{u(t_prime)}
\end{equation}%
with $\mathcal{C}=\cos (\omega _{0}\tau )$ and $\mathcal{S}=\sin (\omega
_{0}\tau )$. So, the back-action of the quantum dot produces the dependence
on $\dot{u}$, $F_{\mathrm{q}}=F_{\mathrm{q}}\left( u,\dot{u}\right) $, in
the form%
\begin{equation}
F_{\mathrm{q}}(t)=F_{\mathrm{q}0}+\Xi \left[ \mathcal{C}u(t)-\omega _{0}^{-1}%
\mathcal{S}\dot{u}(t)\right] .  \label{f_u_and_deriv}
\end{equation}%
This together with Eqs.~(\ref{small_oscs}, \ref{Dw}, \ref{DQ}) provides
expressions for the effective frequency and the quality factor shifts:%
\begin{eqnarray}
\frac{\Delta \omega }{\omega _{0}} &=&-\frac{\mathcal{C}}{2m\omega _{0}^{2}}%
\;\Xi ,  \label{D_w} \\
\frac{\Delta Q}{Q_{0}} &=&-\frac{\mathcal{S}Q_{0}}{m\omega _{0}^{2}}\;\Xi .
\label{D_Q}
\end{eqnarray}%
From these, it follows that the quality factor changes $\Delta Q$ are
directly related to the changes in the frequency shift $\Delta \omega $,
i.e. $\Delta Q\propto \Delta \omega $. Moreover, their ratio quantifies the
delay measure$~\omega _{0}\tau $%
\begin{equation}
\tan (\omega _{0}\tau )=\frac{1}{2Q_{0}}\frac{\Delta Q/Q_{0}}{\Delta \omega
/\omega _{0}}.  \label{tan}
\end{equation}

Note that if the changes of the quality factor $\Delta Q$ are not small, one
should use Eq.~(\ref{Q}) instead of Eq.~(\ref{DQ}). In any case, the quality
factor changes can be termed as the \textquotedblleft
Sisyphus\textquotedblright\ addition to the quality factor~[%
\onlinecite{Skinner10}] as follows%
\begin{equation}
\frac{1}{Q_{\mathrm{eff}}}=\frac{1}{Q_{0}}+\frac{1}{Q_{\text{Sis}}}\text{, \
\ \ \ \ \ \ \ \ }Q_{\text{Sis}}^{-1}=\frac{\mathcal{S}}{m\omega _{0}^{2}}%
\;\Xi .  \label{Q_Sis}
\end{equation}%
Positive values of $Q_{\text{Sis}}$ give rise to damping, while negative
values result in amplification, which is the precursor of lasing~[%
\onlinecite{Grajcar08}]. Here a special case is when $Q_{\text{Sis}%
}\rightarrow -Q_{0}$: this corresponds to the theoretical lasing limit~[%
\onlinecite{Skinner10, Ella14}], in which the regime of self-sustaining
oscillations is realized.

The delayed response can also be related to the work done on the resonator
by the quantum system, QD~[\onlinecite{Clerk05, ShevchenkoIO12}]. The
respective energy transfer during one period is given by
\begin{equation}
W=\oint duF_{\mathrm{q}}=\int\limits_{0}^{2\pi /\omega _{0}}\!\!\!dt\;F_{%
\mathrm{q}}\;\frac{du}{dt}=-\mathcal{S}\pi v^{2}\;\Xi ,  \label{W}
\end{equation}%
which is proportional to the quality factor changes:%
\begin{equation}
\frac{W}{W_{0}}=\frac{\Delta Q}{Q_{0}},  \label{WW}
\end{equation}%
where the normalizing factor is $W_{0}=\pi m\omega _{0}^{2}v^{2}/Q_{0}$.
Note that for the driven resonant oscillations $v=F_{\mathrm{p}%
}Q_{0}/m\omega _{0}^{2}$. Therefore, the positive or negative shift in the
quality factor, i.e. the amplification or damping of the NR oscillations, is
related to the respective work done by the QD. Similar processes have been
described as Sisyphus amplification and cooling of the NR~[%
\onlinecite{Grajcar08, Nori08}]. For further discussion see also Appendices
B and C. Note also that such periodic processes are similar to quantum
thermodynamic cycles, which can be used as quantum heat engines [%
\onlinecite{Grajcar08, Quan07,
Chotorlishvili11}].

\subsection{Quantum dot response}

Let us now explicitly define the back-action force $F_{\mathrm{q}}$ for the
system presented in Fig.~\ref{Fig:scheme}. It is assumed that the mechanical
frequency $\omega _{0}$ is much smaller than the QD tunnelling rate $\Gamma $%
, hence the NR sees the QD charge averaged over many stochastic tunneling
events [\onlinecite{Meerwaldt12}]. Supposing this, the averaged QD charge is
given by
\begin{eqnarray}
e\left\langle n\right\rangle &=&C_{\Sigma }\,V_{\mathrm{I}}+e\,n_{\mathrm{g}%
},  \label{charge} \\
n_{\mathrm{g}} &=&-\frac{1}{e}\left[ C_{2}V_{\mathrm{SD}}+C_{\mathrm{g}}V_{%
\mathrm{g}}+C_{\mathrm{NR}}V_{\mathrm{NR}}(t)\right] .
\end{eqnarray}%
It follows that $V_{\mathrm{I}}=e(\left\langle n\right\rangle -n_{\mathrm{g}%
})/C_{\Sigma }$. Here it is assumed that the NR is biased by a dc plus an ac
voltage: $V_{\mathrm{NR}}(t)=V_{\mathrm{NR}}+V_{\mathrm{A}}\sin \omega _{0}t$%
. Then the electrostatic force becomes%
\begin{eqnarray}
F &=&\frac{\partial }{\partial u}\frac{C_{\mathrm{NR}}(u)\left[ V_{\mathrm{NR%
}}(t)-V_{\mathrm{I}}(u)\right] ^{2}}{2}\approx  \label{F} \\
&\approx &\frac{1}{2}\frac{\partial }{\partial u}C_{\mathrm{NR}}(u)\left[ V_{%
\mathrm{NR}}^{2}+2V_{\mathrm{NR}}\left( V_{\mathrm{A}}\sin \omega _{0}t-V_{%
\mathrm{I}}(u)\right) \right] .  \notag
\end{eqnarray}%
Expanding as a Taylor series to second order we obtain%
\begin{eqnarray}
C_{\mathrm{NR}}(u) &\approx &C_{\mathrm{NR}}(0)+\left. \frac{dC_{\mathrm{NR}}%
}{du}\right\vert _{0}u+\left. \frac{d^{2}C_{\mathrm{NR}}}{du^{2}}\right\vert
_{0}\frac{u^{2}}{2}\equiv  \notag \\
&\equiv &C_{\mathrm{NR}}\left( 1+\frac{u}{\xi }+\frac{u^{2}}{2\lambda }%
\right) ,  \label{CNR(u)}
\end{eqnarray}%
and similarly for $\left\langle n\right\rangle $ and $n_{\mathrm{g}}$. The
second term in the r.h.s. of Eq.~(\ref{F}) results in the periodical
driving, $F_{\mathrm{p}}\sin \omega _{0}t$, with $F_{\mathrm{p}}=V_{\mathrm{A%
}}V_{\mathrm{NR}}C_{\mathrm{NR}}/\xi $. Then keeping only the terms defined
by the QD state, we obtain Eq.~(\ref{F_with_Ksi}) with
\begin{equation}
\Xi =\frac{2E_{C}}{\xi ^{2}}\;n_{\mathrm{NR}}^{3}\left( \frac{%
d^{2}\!\left\langle n\right\rangle }{dn_{\mathrm{g}}^{2}}+\frac{2\alpha }{n_{%
\mathrm{NR}}}\frac{d\left\langle n\right\rangle }{dn_{\mathrm{g}}}+\frac{%
\left\langle n\right\rangle -n_{\mathrm{g}}}{n_{\mathrm{NR}}^{2}}\frac{\xi
^{2}}{\lambda }\right) .  \label{Ksi2}
\end{equation}%
Here $E_{C}=e^{2}/2C_{\Sigma }$, $\alpha =1+\xi ^{2}/2\lambda $, and $n_{%
\mathrm{NR}}=-C_{\mathrm{NR}}V_{\mathrm{NR}}/e.$ For estimations it is
useful to note that for the plane-parallel capacitor with distance $d+u$
between the plates: $\xi =-d$, $\lambda =d^{2}/2$, and $\alpha =2$.

We note in passing that the same results as Eqs.~(\ref{F_with_Ksi},~\ref%
{Ksi2}), can be obtained in terms of the quantum capacitance~[%
\onlinecite{Sillanpaa05, Duty05, ShevchenkoAN12}] by introducing the
effective capacitance
\begin{equation}
C_{\mathrm{eff}}=\partial Q_{\mathrm{NR}}/\partial V_{\mathrm{NR}}=C_{%
\mathrm{geom}}+C_{\mathrm{q}}.
\end{equation}%
The effective capacitance consists of the irrelevant geometric component and
the quantum capacitance,
\begin{equation}
C_{\mathrm{q}}=-\frac{eC_{\mathrm{NR}}}{C_{\mathrm{\Sigma }}}\frac{\partial
\left\langle n\right\rangle }{\partial V_{\mathrm{NR}}}.
\end{equation}%
The force $F_{\mathrm{q}}$ is now given in terms of the effective
capacitance as
\begin{equation}
F_{\mathrm{q}}=\frac{\partial }{\partial u}\frac{C_{\mathrm{eff}}V_{\mathrm{%
NR}}^{2}}{2}.
\end{equation}%
By expanding $C_{\mathrm{NR}}(u)$ and $\left\langle n\right\rangle $ as
series in $u$, we obtain Eqs.~(\ref{F_with_Ksi},~\ref{Ksi2}).

To proceed, we require the QD occupation probability $\left\langle
n\right\rangle $, which depends on the gate voltage via $n_{\mathrm{g}}$.
This is related to the QD conductance, $G(V_{\mathrm{g}})=I/V_{\mathrm{SD}}$%
, as follows~[\onlinecite{Lassagne09, Beenaker91}]%
\begin{equation}
G=-\frac{1}{2}\Gamma \,C_{\mathrm{\Sigma }}\,\frac{d\left\langle
n\right\rangle }{dn_{\mathrm{g}}},  \label{G}
\end{equation}%
where $I$ is the source-drain current and $\Gamma $ is the tunneling rate.
The conductance at low temperature is defined by the transmission, $G=G_{0}%
\mathcal{T}$, with the transmission $\mathcal{T}$ given by the Breit-Wigner
formula~[\onlinecite{VanHouten92}]:%
\begin{equation}
G=G_{0}g\frac{\left( \hbar \Gamma \right) ^{2}}{\left( \hbar \Gamma \right)
^{2}+\left[ 2E_{\mathrm{C}}\left( n_{\mathrm{g}}-n_{\mathrm{g}}^{(0)}\right) %
\right] ^{2}}\equiv \frac{G_{0}g}{1+\left( \varepsilon _{0}/\Delta \right)
^{2}},  \label{B-W}
\end{equation}%
where the Lorentzian curve half-width at half-maximum and its center are
defined by $\hbar \Gamma $ and $n_{\mathrm{g}}^{(0)}$. The formula for the
conductance is valid for small tunneling rates, for $E_{\mathrm{C}}\gg \hbar
\Gamma ,k_{\mathrm{B}}T$. Here, $\Gamma =(\Gamma _{1}+\Gamma _{2})/2$ stands
for the averaged tunneling rate into the left ($\Gamma _{1}$) and right ($%
\Gamma _{2}$) reservoirs; the factor $g=\Gamma _{1}\Gamma _{2}/\Gamma ^{2}$
diminishes the conductance and the current if the rates are not equal. We
have also defined here the tunneling amplitude $\Delta $ and the energy bias
$\varepsilon _{0}$, as follows%
\begin{equation}
\varepsilon _{0}=2E_{\mathrm{C}}\left( n_{\mathrm{g}}-n_{\mathrm{g}%
}^{(0)}\right) \text{.}
\end{equation}%
Then the expression for the source-drain current $I$ reads
\begin{equation}
I=I_{0}\frac{1}{1+\left( \varepsilon _{0}/\Delta \right) ^{2}},\text{ \ \ \
\ \ }I_{0}=V_{\mathrm{SD}}G_{0}g.  \label{I}
\end{equation}

Combined together, Eqs.~(\ref{Ksi2}-\ref{B-W}) define the effective quality
factor shift (\ref{D_Q}). For illustration, we take $\left\vert n_{\mathrm{NR%
}}\right\vert \gg 1$, leaving only the second term in Eq.~(\ref{Ksi2}), to
obtain

\begin{eqnarray}
\Delta Q &=&-Q_{S}\frac{\varepsilon _{0}/\Delta }{\left( 1+\left(
\varepsilon _{0}/\Delta \right) ^{2}\right) ^{2}},  \label{DQ/DS} \\
Q_{S} &=&\mathcal{S}\frac{16\pi Q_{0}^{2}\Delta }{m\omega _{0}^{2}\xi ^{2}}%
\left( \frac{2E_{\mathrm{C}}}{\Delta }\right) ^{3}n_{\mathrm{NR}}^{3}.
\notag
\end{eqnarray}%
Note that the theoretical lasing condition~[\onlinecite{Skinner10}] is when $%
Q_{\text{Sis}}=-Q_{0}$ (see Eq.~(\ref{Q_Sis})) and this is fulfilled in our
case when $\Xi (\varepsilon _{0})=-m\omega _{0}^{2}/\mathcal{S}Q_{0}$.

In addition, from Eqs.~(\ref{G}-\ref{B-W}) we also obtain%
\begin{equation}
\left\langle n\right\rangle =-\frac{2}{\Gamma C_{\Sigma }}\int G(n_{\mathrm{g%
}})\;dn_{\mathrm{g}}=-\frac{1}{\pi }\arctan \left( \frac{\varepsilon _{0}}{%
\Delta }\right) +\frac{1}{2}.  \label{n_avrgd}
\end{equation}%
Then, from the QD Hamiltonian [\onlinecite{Wang13, ShevchenkoAN12}],
\begin{equation}
H=-\frac{1}{2}\left( \varepsilon _{0}\sigma _{z}+\Delta \sigma _{x}\right) ,
\label{Ham}
\end{equation}%
\ we have
\begin{equation}
E_{\pm }=\pm \frac{1}{2}\sqrt{\Delta ^{2}+\varepsilon _{0}^{2}}\text{\textit{%
\ .}}  \label{Epm}
\end{equation}%
The above results, Eqs.~(\ref{I}-\ref{Epm}), are illustrated in Fig.~\ref%
{Fig:I-Q} and discussed below.

\begin{figure}[t]
\includegraphics[width=7.7 cm]{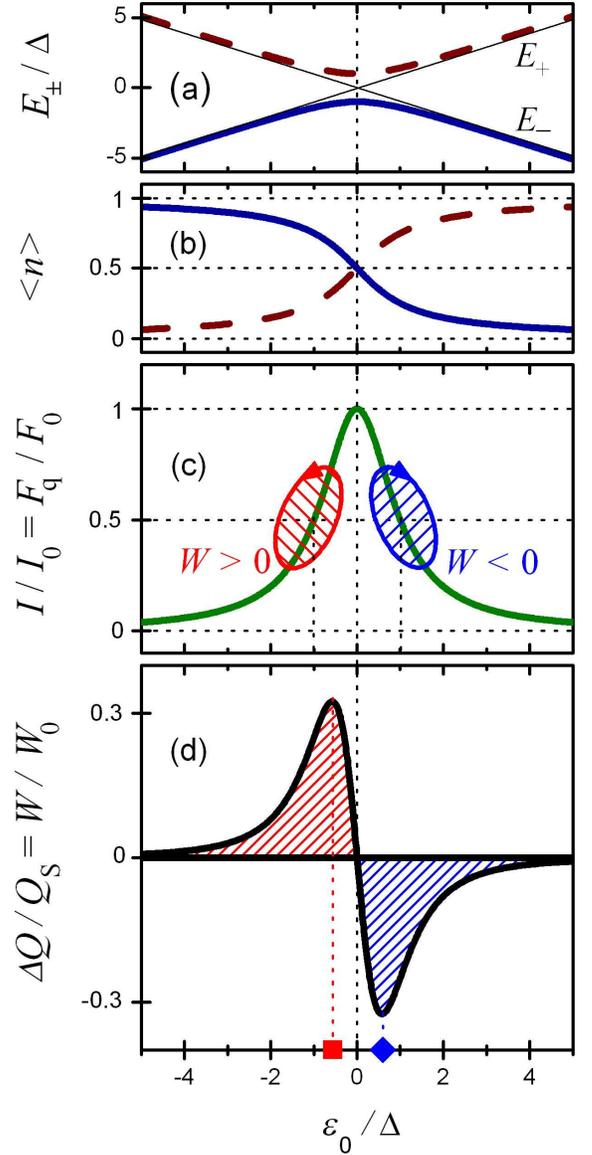}
\caption{(Color online) The gate-voltage offset $\protect\varepsilon _{0}$
dependence of: (a)~the energy levels $E_{\pm }/\Delta $; (b)~average
electron number $\left\langle n\right\rangle $; (c)~QD current $I$ and the
force $F_{\mathrm{q}}$ which influences the NR, where $I/I_{0}=F_{\mathrm{q}%
}/F_{0}$; and (d)~the NR quality-factor changes $\Delta Q$ and the work $W$
on the NR, where $\Delta Q/Q_{\mathrm{S}}=W/W_{0}$. Closed trajectories in
(c) describe the delayed value of the force, and the hatched areas give the
work $W=\protect\oint duF_{\mathrm{q}}$. If there is no delay ($%
T_{1}\rightarrow 0$), then the red and blue hatched ovals in (c) merge with
the solid green curve, which describes the adiabatic evolution. The
source-drain current $I$ in (c) is given by the Lorentzian (green) curve and
the quality factor changes $\Delta Q$ in (d) are defined by its derivative,
of which the maximum and the minimum are indicated by a square and a
rhombus. }
\label{Fig:I-Q}
\end{figure}

\section{Discussion}

Figure~\ref{Fig:I-Q} graphically describes the interaction of the NR and the
QD. The controllable parameter is the offset $\varepsilon _{0}=2E_{\mathrm{C}%
}(n_{\mathrm{g}}-n_{\mathrm{g}}^{(0)})$, which can be influenced by both the
gate voltage $V_{\mathrm{g}}$ and the NR displacement $u$.

The ground- and excited-state energy levels of the QD are plotted in Fig.~%
\ref{Fig:I-Q}(a), while the respective average excessive electron number $%
\left\langle n\right\rangle $ is shown in Fig.~\ref{Fig:I-Q}(b). If the gate
voltage $V_{\mathrm{g}}$ is fixed, the evolution is described by the changes
in the NR displacement $u$. Its influence on the QD is discussed in Appendix
B. Here we concentrate on the back-action.

Figure~\ref{Fig:I-Q}(c) shows the Lorentzian-shaped dependence of the
current $I$ through the QD, given by Eq.~(\ref{I}). We note that the
dependence of the force $F_{\mathrm{q}}$, which influences the NR, is
similar. To demonstrate this, we find the expression of the
displacement-dependent force from Eqs.~(\ref{F_with_Ksi}, \ref{Ksi2}): for
the changes $\Delta u$ we have $\Delta F_{\mathrm{q}}=\Xi \Delta u$. Then,
integrating this and assuming $\left\vert n_{\mathrm{NR}}\right\vert \gg 1$,
we obtain%
\begin{eqnarray}
F_{\mathrm{q}}(u) &=&\frac{2E_{C}n_{\mathrm{NR}}^{2}}{\xi }\left( \frac{%
d\left\langle n\right\rangle }{dn_{\mathrm{g}}}\right) =F_{0}\frac{1}{1+%
\left[ \varepsilon _{0}(u)/\Delta \right] ^{2}},\text{ }  \label{F_q} \\
F_{0} &=&\frac{8gE_{C}^{2}n_{\mathrm{NR}}^{2}}{\pi \xi \Delta }.  \notag
\end{eqnarray}%
This means that $I/I_{0}=F_{\mathrm{q}}/F_{0}$, and Fig.~\ref{Fig:I-Q}(c)
describes both the current $I$ and the force $F_{\mathrm{q}}$. We note that $%
\Delta n_{\mathrm{g}}=n_{\mathrm{NR}}\Delta u/\xi $, and thus $\Delta n_{%
\mathrm{g}}$ and $\Delta u$ have opposite signs for negative $V_{\mathrm{NR}%
} $.

Alongside the discussion in Appendix A, the closed oval trajectories in Fig.~%
\ref{Fig:I-Q}(c) indicate the essence of the delayed-response method of
analysis of the NR-QD system; see also Ref.~[\onlinecite{Poot12}]. The
periodic evolution of the NR displacement $u$ results in the periodic
sweeping the bias $\varepsilon _{0}(u)$ about its value at $u=0$, defined by
the gate voltage $V_{\mathrm{g}}$. In Fig.~\ref{Fig:I-Q}(c) we demonstrate
two such situations with $\varepsilon _{0}(u=0)=\pm \Delta $ as examples.
The points on the elliptical curves give the value of the force $F_{\mathrm{q%
}}$ for some previous time $\widetilde{t}=t-\tau $. In particular, if there
is no delay ($\tau \rightarrow 0$), these ovals in Fig.~\ref{Fig:I-Q}(c)
shrink to the solid green curve, which describes the adiabatic evolution. In
contrast to this, the back-action with delay results in two types of
trajectories, shown in Fig.~\ref{Fig:I-Q}(c), of which the non-zero area
gives the work done by the drivings via the QD on the NR, $W=\oint duF_{%
\mathrm{q}}\gtrless 0$; see Eq.~(\ref{W}). One can see from this geometric
interpretation that the back-action effect is maximum when $\tau =T_{0}/4$,
when the ovals tend to circles and the weight of the respective quadrature
in Eq.~(\ref{u(t_prime)}) becomes maximal, at $\mathcal{S}=1$.

Finally, figure~\ref{Fig:I-Q}(d) displays the gate-voltage offset dependence
of the quality factor changes $\Delta Q$. We emphasize that, in agreement
with Eqs.~(\ref{D_w}, \ref{D_Q}, \ref{WW}), we have
\begin{equation}
\Delta Q\propto \Delta \omega \text{ \ \ \ and \ \ }\Delta Q\propto W,
\end{equation}%
which means that Fig.~\ref{Fig:I-Q}(d) can also be interpreted (up to a
normalizing factor) as the gate-voltage dependence of the frequency shift $%
\Delta \omega $\ and the work $W$ done by the QD on the NR.

\begin{figure}[t]
\includegraphics[width=8 cm]{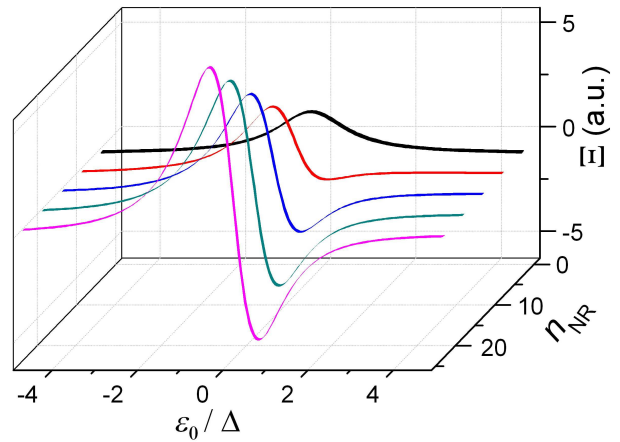}
\caption{(Color online) The bias $\protect\varepsilon _{0}$ dependence of
the response function $\Xi $ for several values of the NR voltage, $n_{%
\mathrm{NR}}=-C_{\mathrm{NR}}V_{\mathrm{NR}}/e$. For $n_{\mathrm{NR}}\sim 1$
the response is described by a peak at $\protect\varepsilon _{0}=0$, while
for $\left\vert n_{\mathrm{NR}}\right\vert \gg 1$ there are both an increase
and a decrease of the function, which relates to the quality factor changes $%
\Delta Q$, as demonstrated in Fig.~\protect\ref{Fig:I-Q}(d).}
\label{Fig:Ksi}
\end{figure}

In Fig.~\ref{Fig:Ksi} the response function $\Xi $ is plotted for several
values of the NR voltage, $n_{\mathrm{NR}}=-C_{\mathrm{NR}}V_{\mathrm{NR}}/e$%
. For this we used Eq.~(\ref{Ksi2}) without assuming $\left\vert n_{\mathrm{%
NR}}\right\vert \gg 1$. Recall that the response function $\Xi $ is the
function which defines the quality factor changes, Eq.~(\ref{D_Q}). Figure~%
\ref{Fig:Ksi} demonstrates how for small values of $n_{\mathrm{NR}}$ the
response is described by the first term in Eq.~(\ref{Ksi2}), while for large
$\left\vert n_{\mathrm{NR}}\right\vert \gg 1$, it is defined by the second
term. In this way the first term describes only positive values of the
response, while the second term can be both positive and negative and can
result in respective changes of the quality factor; see also in Ref.~[%
\onlinecite{Bode12}].

\section{Conclusions}

We have presented a quasiclassical theory for the \textquotedblleft quantum
dot -- nanomechanical resonator\textquotedblright\ system using a
phenomenological delayed-response method. This method is a useful and
intuitive tool for the description of a coalesced system, where a
slowly-driven subsystem (resonator) is coupled to a quantum subsystem. The
relaxation of the latter results in the delayed back-action. The advantage
of this method over the use of a master equation is in the detachment of the
dynamics of the two subsystems. The delayed response is included via the
simple substitution $t\rightarrow \widetilde{t}=t-\tau $. This means that
the back-action force $F_{\mathrm{q}}$ is time-delayed via the displacement $%
u$ by the characteristic relaxation time $\tau $: $F_{\mathrm{q}}(t)=F_{%
\mathrm{q}}\left[ u(t-\tau )\right] $.

Our theory describes the increase and decrease of the NR quality factor due
to the phase-shifted back-action force. This can be interpreted as Sisyphus
cooling and amplification of the NR oscillations. This approach can be
useful for the description and interpretation of experiments, such as those
in Refs.~[\onlinecite{Grajcar08, OkazakiISNTT13}].

\begin{acknowledgments}
We are grateful to Y. Okazaki and H. Yamaguchi for stimulating discussions
of their experimental results~[\onlinecite{OkazakiISNTT13}]. We thank Neill
Lambert for advice and discussions and Sophia Lloyd for carefully reading
the manuscript. This research is partially supported by the RIKEN iTHES
Project, MURI Center for Dynamic Magneto-Optics, a Grant-in-Aid for
Scientific Research (S), DKNII (project no. F52.2 /009), and the NAS of
Ukraine (project no.~4/14-NANO).
\end{acknowledgments}

\appendix

\section{Justification of the delayed-response method}

Here we present the justification for the delayed-response method, which was
formulated in the introduction and applied afterwards. Consider the force,
which influences a resonator, to be exponentially decaying,%
\begin{equation}
F_{\mathrm{q}}(t)=F_{\mathrm{q}0}+\left[ F_{\mathrm{q}}(t_{0})-F_{\mathrm{q}%
0}\right] \exp \left( -\frac{t-t_{0}}{T_{1}}\right) .  \label{Fq(t)}
\end{equation}%
This is given at the initial moment, $t=t_{0}$, by $F_{\mathrm{q}}(t_{0})$
and tends to an equilibrium value $F_{\mathrm{q}0}$ with increasing time.
The force enters the r.h.s.~of the resonator motion equation, Eq.~(\ref{(1)}%
). Consider the case of small retardation parameter,%
\begin{equation}
\omega _{0}T_{1}\ll 1,  \label{limiting_case}
\end{equation}%
which means that the relaxation happens fast in respect to the resonator
period $T_{0}=2\pi /\omega _{0}$. It is then reasonable to average Eq.~(\ref%
{(1)}) during the time interval $\Delta t\sim T_{1}$. According to the
assumption, during this interval one can neglect the changes in the
resonator evolution, leaving the l.h.s.~of Eq.~(\ref{(1)}) unaffected. Next
we assume the linear displacement dependence
\begin{equation}
F_{\mathrm{q}}(t)-F_{\mathrm{q}0}=\Xi u(t),\text{ \ \ \ \ \ \ }\Xi =\left.
\frac{dF_{\mathrm{q}}}{du}\right\vert _{u=0},
\end{equation}%
and obtain for the averaged force%
\begin{eqnarray}
\overline{F_{\mathrm{q}}}(t) &\equiv &\frac{1}{\Delta t}\int\limits_{t-%
\Delta t}^{t}dt^{\prime }F_{\mathrm{q}}(t^{\prime }) \\
&=&F_{\mathrm{q}0}+\frac{1}{\Delta t}\int\limits_{t-\Delta t}^{t}dt^{\prime }%
\left[ F_{\mathrm{q}}(t-\Delta t)-F_{\mathrm{q}0}\right] e^{-\frac{t^{\prime
}-(t-\Delta t)}{T_{1}}}  \notag \\
&=&F_{\mathrm{q}0}+\Xi u(t-\Delta t)\cdot f\left( \frac{T_{1}}{\Delta t}%
\right) ,  \notag
\end{eqnarray}%
where $f(x)=x\left( 1-e^{-1/x}\right) $. Then, choosing $\Delta t=T_{1}$ and
neglecting distinction of $f(1)$ from unity, one obtains that the delayed
force enters the equation of motion of the resonator,%
\begin{equation}
\overline{F_{\mathrm{q}}}(t)=F_{\mathrm{q}0}+\Xi u(t-T_{1}).
\label{DRM_force}
\end{equation}%
This justifies the delayed-response approximation and results in the
velocity-dependence of the force%
\begin{equation}
\overline{F_{\mathrm{q}}}(t)=F_{\mathrm{q}}(u,\dot{u}),
\end{equation}%
as it was discussed in the main text, see Eqs.~(\ref{u(t_prime)}-\ref%
{f_u_and_deriv}).

Here we note that our Eq.~(\ref{DRM_force}) gives the result consistent with
those used in Refs.~[\onlinecite{Metzger04, Clerk05}]. For comparison we
rewrite here the respective averaged forces in our notations:%
\begin{eqnarray}
\text{\negthinspace \negthinspace \negthinspace \negthinspace \negthinspace
\negthinspace \negthinspace \negthinspace \negthinspace \negthinspace
\negthinspace \negthinspace \lbrack \onlinecite{Metzger04}]\negthinspace
\negthinspace \negthinspace } &:&\overline{F_{\mathrm{q}}}(t)\text{%
\negthinspace }=\text{\negthinspace \negthinspace \negthinspace }%
\int\limits_{0}^{t}\text{\negthinspace }dt^{\prime }\frac{dF_{\mathrm{q}}%
\text{\negthinspace }\left[ u(t^{\prime })\right] }{dt^{\prime }}\text{%
\negthinspace }\left( \text{\negthinspace }1\text{\negthinspace }-\text{%
\negthinspace }\exp \text{\negthinspace \negthinspace }\left( \text{%
\negthinspace \negthinspace }-\frac{t-t^{\prime }}{T_{1}}\text{\negthinspace
}\right) \text{\negthinspace \negthinspace }\right) \text{\negthinspace ,}
\label{Metzger} \\
\text{\negthinspace \negthinspace \negthinspace \negthinspace \negthinspace
\negthinspace \negthinspace \negthinspace \negthinspace \negthinspace
\negthinspace \negthinspace \lbrack \onlinecite{Clerk05}]\negthinspace
\negthinspace \negthinspace } &:&\ \overline{F_{\mathrm{q}}}(t)=\frac{d%
\overline{F_{\mathrm{q}}}}{dt}\int\limits_{-\infty }^{t}dt^{\prime }\exp
\left( -\frac{t-t^{\prime }}{T_{1}}\right) u(t^{\prime })\text{.}
\label{Clerk}
\end{eqnarray}%
One can check that the three equations, Eqs.~(\ref{DRM_force}, \ref{Metzger}%
, and \ref{Clerk}), result for the steady-state oscillations in the same
Eq.~(\ref{f_u_and_deriv}) in the limiting case of Eq.~(\ref{limiting_case}).

Consider now the origin of Eq.~(\ref{Fq(t)}) in our problem of the
qubit-resonator system. The system is described by the equation for the
resonator displacement $u(t)$, Eq.~(\ref{(1)}), plus the Bloch equations for
the reduced qubit density matrix $\rho =\frac{1}{2}(1+X\sigma _{x}+Y\sigma
_{y}+Z\sigma _{z})$ with the relaxation times $T_{1,2}$,%
\begin{eqnarray}
\dot{X} &=&\left( \frac{\Delta E}{\hbar }+\frac{\varepsilon _{0}}{\Delta }%
\beta u\right) Y-\frac{X}{T_{2}},  \label{Bloch} \\
\dot{Y} &=&-\left( \frac{\Delta E}{\hbar }+\frac{\varepsilon _{0}}{\Delta }%
\beta u\right) X-\beta uZ-\frac{Y}{T_{2}},  \notag \\
\dot{Z} &=&\beta uY-\frac{Z-Z^{(0)}}{T_{1}},  \notag
\end{eqnarray}%
where%
\begin{equation}
\beta =\frac{2E_{C}n_{\mathrm{NR}}}{\hbar \xi }\frac{\Delta }{\Delta E}\text{%
, \ \ \ }Z^{(0)}=\tanh \frac{\Delta E}{2k_{\mathrm{B}}T}\text{.}
\end{equation}%
Here $Z^{(0)}$ corresponds to the equilibrium value at nonzero temperature $%
T $. Then, if the coupling $\beta $ between the resonator and the qubit is
small and/or the oscillations $u(t)$ are small, one can neglect the first
term in the equation for $Z(t)$, Eq.~(\ref{Bloch}). This results in the
exponential dependence, as in Eq.~(\ref{Fq(t)}).

\begin{figure}[b]
\includegraphics[width=6 cm]{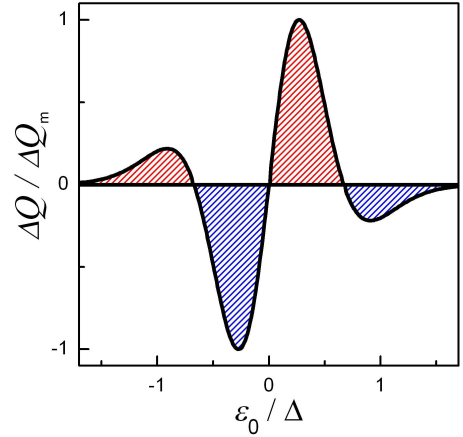}
\caption{(Color online) Normalized quality factor shift $\Delta Q$ as the
function of the energy bias $\protect\varepsilon _{0}$ calculated with Eq.~(%
\protect\ref{DQ_with_T}).}
\label{Fig:Q_T}
\end{figure}

To be more specific, consider a QD with the Hamiltonian written in the
charge representation in the two-level approximation in Eq.~(\ref{Ham}).
Relating the charge and eigen-bases, we have%
\begin{equation}
\left\langle n\right\rangle =P_{-}\left\langle n\right\rangle
_{-}+P_{+}\left\langle n\right\rangle _{+}=\left\langle n\right\rangle
_{-}+P_{+}\left( \left\langle n\right\rangle _{+}-\left\langle
n\right\rangle _{-}\right) ,  \label{n}
\end{equation}%
where the level occupation probabilities are $P_{\pm }=\frac{1}{2}\left(
1\mp Z\right) $ and we defined the coefficients
\begin{equation}
\left\langle n\right\rangle _{\pm }=\frac{1}{2}\left( 1\pm \frac{\varepsilon
_{0}}{\Delta E}\right) ,  \label{n_pm}
\end{equation}%
then $\left\langle n\right\rangle _{+}-\left\langle n\right\rangle
_{-}=\varepsilon _{0}/\Delta E$. We note that in the absence of excitation, $%
P_{+}=0$, we have $\left\langle n\right\rangle =\left\langle n\right\rangle
_{-}$, which is in good agreement with the assumption of the Breit-Wigner
tunneling, cf. Eqs. (\ref{B-W},~\ref{n_avrgd}). In the other case, in
thermal equilibrium, from Eq.~(\ref{n}) we have [\onlinecite{Wang13}]%
\begin{equation}
\left\langle n\right\rangle =\frac{1}{2}-\frac{\varepsilon _{0}}{2\Delta E}%
\tanh \frac{\Delta E}{2k_{\mathrm{B}}T}.
\end{equation}

In this picture, the delayed response is related to the nonzero upper level
occupation, which is the latter term in Eq.~(\ref{n}), rather than to the
ground-state average number $\left\langle n\right\rangle _{-}$. With this
note, combining the equations above, we obtain the formula for the quality
factor phase shift, which for $n_{\mathrm{NR}}\gg 1$ reads%
\begin{equation}
\Delta Q\approx -\mathcal{S}\frac{4Q_{0}^{2}E_{\mathrm{C}}^{3}n_{\mathrm{NR}%
}^{3}}{m\omega _{0}^{2}\xi ^{2}}\frac{d^{2}}{d\varepsilon _{0}^{2}}\left[
\frac{\varepsilon _{0}}{\Delta E}\left( 1-\tanh \frac{\Delta E}{2k_{\mathrm{B%
}}T}\right) \right] ,  \label{DQ_with_T}
\end{equation}%
where the retardation parameter is defined by the relaxation time, $\mathcal{%
S}=\sin \left( \omega _{0}T_{1}\right) $.

We illustrate the result, Eq.~(\ref{DQ_with_T}), in Fig.~\ref{Fig:Q_T},
where the quality factor shift $\Delta Q$ is normalized to its maximal value
$\Delta Q_{m}$ and is plotted as the function of the energy bias $%
\varepsilon _{0}=\varepsilon _{0}(V_{\mathrm{g}})$ for $k_{\mathrm{B}%
}T=0.1\Delta $. The figure demonstrates the amplification and attenuation of
the NR oscillations. These can be interpreted in terms of the Sisyphus
cycles, which we detail below. Here we emphasize that the important feature
of the process is the double-amplification/attenuation structure,
demonstrated in Fig.~\ref{Fig:Q_T}. This may be useful in analyzing the
experimental results such as those detailed in Ref.~[%
\onlinecite{OkazakiISNTT13}].

\section{Sisyphus cycles for the nanoelectromechanical system}

\begin{figure}[t]
\includegraphics[width=8 cm]{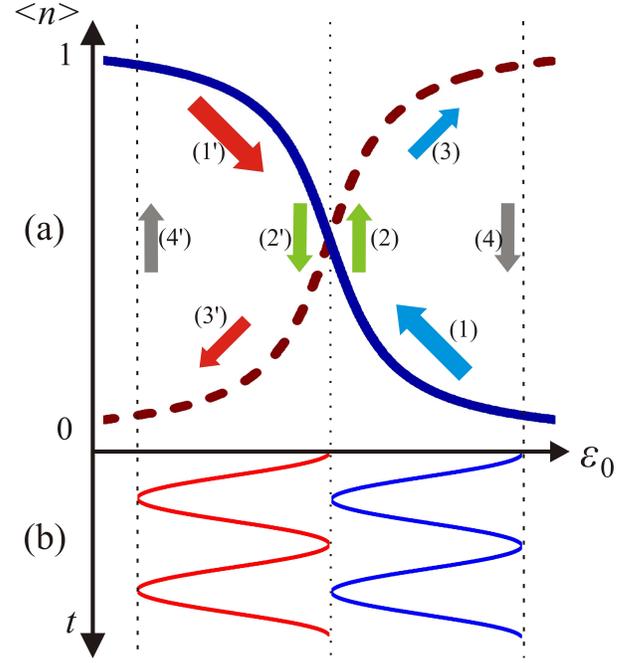}
\caption{(Color online) (a) Average excessive electron number on the QD $%
\left\langle n\right\rangle $ as a function of the bias $\protect\varepsilon %
_{0}$ and (b) changes of the bias due to the periodic evolution of the NR.
The right and left halves of the graph illustrate the cycles in which
resonator changes the average number of electrons from 0 to 1 and vice
versa. See text for a detailed description of these cycles.}
\label{Fig:Sisyphus}
\end{figure}

In the main text we were principally interested in the back-action effect.
In particular, Fig.~\ref{Fig:I-Q}(c) shows the work over the NR during one
period. Here we consider this evolution as seen by the QD. For this goal, in
Fig.~\ref{Fig:Sisyphus} we consider the average excessive electron number $%
\left\langle n\right\rangle $ versus the bias $\varepsilon _{0}$. These are
the same curves as in Fig.~\ref{Fig:I-Q}(b), plotted with Eq.~(\ref{n}),
where the solid line corresponds to the ground state and the dashed one to
the excited state. We consider slow periodic changes of the NR displacement,
which correspond to changing the bias, see Fig.~\ref{Fig:Sisyphus}(b). Note
that for illustration we consider the number of electrons and not the energy
levels since we have an open driven system in which energy changes of one
subsystem should not be equal to minus the energy changes in another
subsystem.

In the right and left halves of Fig.~\ref{Fig:Sisyphus} we consider two
cases of positive and negative offsets. The amplitude of the oscillations is
chosen to be twice the offset, so that the resonator drives the two-level
system (TLS) between the point of energy level quasi-intersection (at $%
\varepsilon _{0}=0$) and the point removed from it; see also Fig.~\ref%
{Fig:I-Q}(a,b). We assume that the region where the energy levels are curved
(i.e., experience avoided-level crossing) plays the role of a 50/50-
beam-splitter. This means that after going out of this region, the TLS
levels are equally populated. Here we assume that the characteristic
relaxation time $T_{1}$ is longer than the time of passing this region.
Moreover, we assume that it is of the order of the driving period, namely, $%
T_{1}\sim T_{0}/4=\pi /2\omega _{0}$.

Then, the overall dynamics of the TLS can be stroboscopically split in four
intervals. Consider first the right part of Fig.~\ref{Fig:Sisyphus}. (1) The
resonator drives the TLS uphill along the ground state, $\left\langle
n\right\rangle $ changes from 0 to 1/2. (2) In the region of the
avoided-level crossing, the two energy levels become equally populated; and
then we monitor the upper-level evolution. (3) Again the resonator drives
the TLS uphill, until it relaxes during the fourth evolution stage. In this
cycle the resonator does work such that it changes $\left\langle
n\right\rangle $ from 0 to 1, while the relaxation does vice versa. In
contrast, in the inverted cycle, $(1^{\prime }-4^{\prime })$, shown in the
left part of Fig.~\ref{Fig:Sisyphus}, the resonator does work changing $%
\left\langle n\right\rangle $ from 1 to 0.

The beam-splitting can be created in several ways. (i) This can be created
by means of non-adiabatic Landau-Zener transitions between the energy
levels~[\onlinecite{Shevchenko10, Petta10, Sun10, Satanin12, Stehlik12}].
(ii) The 50/50- beam-splitting can be created by resonantly driving the TLS
as in Ref.~[\onlinecite{Grajcar08}]. (iii) Alternatively, the non-zero
upper-level occupation can be created by the thermal excitation, which is
essential when the temperature is comparable with the energy-level
separation, as it was considered in the previous section.

\section{Sisyphus cycles described with the delayed-response theory}

The equations for the source-drain current $I$ and the changes of the NR
quality factor $\Delta Q$ can be rewritten as follows:%
\begin{eqnarray}
\frac{I}{I_{0}} &=&\frac{1}{1+(\varepsilon _{0}/\Delta )^{2}},  \label{ISD}
\\
\Delta Q &\propto &\mathcal{S}\left( I+a\frac{dI}{d\varepsilon _{0}}\right) ,
\label{DQ_prop}
\end{eqnarray}%
where $a=E_{\mathrm{C}}n_{\mathrm{NR}}/\alpha $. The former equation was
illustrated in Fig.~\ref{Fig:I-Q}(c). The latter equation was illustrated in
Fig. \ref{Fig:Ksi}. The deep analogy with the Sisyphus cycles for the flux
qubit-\textit{LC} resonator system~[\onlinecite{Grajcar08}], mentioned
earlier in the text, can be further justified by writing down analogous
equations for this system. So, following Ref.~[\onlinecite{ShevchenkoIO12}],
we consider now the driven flux qubit with the Hamiltonian
\begin{equation}
H=-\frac{\varepsilon _{0}+A\sin \omega _{\mathrm{d}}t}{2}\sigma _{z}-\frac{%
\Delta }{2}\sigma _{x}.
\end{equation}%
In this case, the averaged current in the flux qubit is~[%
\onlinecite{ShevchenkoIO12}] $I_{\mathrm{qb}}=I_{\mathrm{p}}\frac{%
\varepsilon _{0}}{\Delta E}(2P_{+}-1)$, where $I_{\mathrm{p}}$ is the flux
qubit persistent current and the averaged upper level occupation probability
is given by the Lorentzian%
\begin{equation}
P_{+}=\frac{1}{2}\frac{1}{1+(\delta \varepsilon _{0}/\hbar \Omega )^{2}},
\label{Pp}
\end{equation}%
with $\delta \varepsilon _{0}=\varepsilon _{0}-\hbar \omega _{\mathrm{d}}$
and $\hbar \Omega =\Delta A/2\hbar \omega _{\mathrm{d}}$. Then for the
changes of the quality factor $\Delta Q$ of the\textit{\ LC} resonator one
can obtain~[\onlinecite{ShevchenkoIO12}]%
\begin{equation}
\Delta Q\;\propto \;\mathcal{S}\left( P_{+}+b\ \frac{dP_{+}}{d\varepsilon
_{0}}\right) ,  \label{DQ_prop2}
\end{equation}%
where $b=\Delta E^{2}\,\varepsilon _{0}/\Delta ^{2}$. This equation is fully
analogous to Eq.~(\ref{DQ_prop}); it is proportional to the lagging
parameter $\mathcal{S}$ (which is zero at $T_{1}=0$) and contains two
competing terms: the Lorentzian and its derivative; the latter being the
alteration of a peak and a dip. It is this latter term (when it is dominant)
that describes the Sisyphus amplification and cooling, respectively~[%
\onlinecite{Grajcar08}].

\section{Derivation of Eq.~(20)}

The most essential appendices are the other ones. This final appendix is
more technical; here we present a more detailed derivation of Eq.~(\ref{Ksi2}%
) in the main text, in addition to the theory in Sec.~II C. There we
considered the averaged QD charge given by the sum of the charges on the
plates of the capacitors, which create the QD:
\begin{equation}
e\left\langle n\right\rangle =\sum\limits_{i}C_{i}\left( V_{\mathrm{I}%
}-V_{i}\right) =V_{\mathrm{I}}\sum\limits_{i}C_{i}-\sum\limits_{i}C_{i}V_{i}%
\equiv C_{\Sigma }\,V_{\mathrm{I}}+e\,n_{\mathrm{g}},
\end{equation}%
where $V_{\mathrm{I}}$ is the QD potential and $V_{i}$ is the voltage
applied to the $i$-th capacitance $C_{i}$. Then the electrostatic force,
Eq.~(\ref{F}), becomes%
\begin{eqnarray}
F &\approx &\frac{1}{2}\frac{d}{du}C_{\mathrm{NR}}(u)\left[ V_{\mathrm{NR}%
}^{2}+2V_{\mathrm{NR}}\left( V_{\mathrm{A}}\sin \omega _{0}t-V_{\mathrm{I}%
}(u)\right) \right] \equiv   \notag \\
&\equiv &F_{\mathrm{q}}+F_{\mathrm{p}}\sin \omega _{0}t,  \label{Fb}
\end{eqnarray}%
where it was assumed that $V_{\mathrm{NR}}\gg V_{\mathrm{A}},V_{\mathrm{I}}$%
. Expanding as a Taylor series to second order we obtain Eq.~(\ref{CNR(u)}).
The second term in Eq.~(\ref{Fb}) results in the periodic driving, $F_{%
\mathrm{p}}\sin \omega _{0}t$. Note that there is also an explicit time
dependence in the third term, where $V_{\mathrm{NR}}(t)$ also enters in $V_{%
\mathrm{I}}$ (via $n_{\mathrm{g}}$); then in addition to the second term
there is small term which can be neglected:%
\begin{equation}
V_{\mathrm{A}}\sin \omega _{0}t-\frac{C_{\mathrm{NR}}}{C_{\Sigma }}V_{%
\mathrm{A}}\sin \omega _{0}t\approx V_{\mathrm{A}}\sin \omega _{0}t\text{, }
\end{equation}%
assuming $C_{\mathrm{NR}}\ll C_{\Sigma }$. Now we have%
\begin{equation}
F_{\mathrm{q}}=\frac{V_{\mathrm{NR}}^{2}}{2}\frac{d}{du}C_{\mathrm{NR}}(u)-%
\frac{eV_{\mathrm{NR}}}{C_{\Sigma }}\frac{d}{du}C_{\mathrm{NR}}(u)\left(
\left\langle n\right\rangle -n_{\mathrm{g}}\right) .  \label{Fq}
\end{equation}%
The displacement-dependence in $C_{\Sigma }$ can be neglected for $C_{%
\mathrm{NR}}\ll C_{\Sigma }$:%
\begin{equation}
C_{\Sigma }(u)=C_{\Sigma }^{(0)}\left( 1+\frac{C_{\mathrm{NR}}^{(0)}}{%
C_{\Sigma }^{(0)}}\frac{u}{\xi }\right) \approx C_{\Sigma }^{(0)}\equiv
C_{\Sigma }\text{.}
\end{equation}%
Note that here and below, for brevity, we omit the superscript (0): $%
C_{\Sigma }^{(0)}=C_{\Sigma }(u=0)\equiv C_{\Sigma }$. Other values are
expanded, making use of Eq.~(\ref{CNR(u)}) and also neglecting the explicit
time dependence in $n_{\mathrm{g}}$, as we noted above:%
\begin{eqnarray}
n_{\mathrm{g}}(u) &=&-\frac{1}{e}\left[ C_{2}V_{\mathrm{SD}}+C_{\mathrm{g}%
}V_{\mathrm{g}}+C_{\mathrm{NR}}(u)V_{\mathrm{NR}}\right] \approx
\label{ng_new} \\
&\approx &n_{\mathrm{g}0}+n_{\mathrm{NR}}\left( \frac{u}{\xi }+\frac{u^{2}}{%
2\lambda }\right) ,  \notag
\end{eqnarray}%
\begin{equation}
\left\langle n\right\rangle (u)\approx \left. \left\langle n\right\rangle
\right\vert _{0}+\left. \frac{d\!\left\langle n\right\rangle }{du}%
\right\vert _{0}u+\left. \frac{d^{2}\!\left\langle n\right\rangle }{du^{2}}%
\right\vert _{0}\frac{u^{2}}{2}.
\end{equation}%
It is convenient to change the derivative from $u$ to $n_{\mathrm{g}}$,
making use of Eq.~(\ref{ng_new}):%
\begin{equation}
\left. \frac{d\!\left\langle n\right\rangle }{du}\right\vert _{0}=\left.
\frac{d\!\left\langle n\right\rangle }{dn_{\mathrm{g}}}\frac{dn_{\mathrm{g}}%
}{du}\right\vert _{0}=\frac{d\!\left\langle n\right\rangle }{dn_{\mathrm{g}}}%
\frac{n_{\mathrm{NR}}}{\xi };
\end{equation}%
\begin{eqnarray}
\left. \frac{d^{2}\!\left\langle n\right\rangle }{du^{2}}\right\vert _{0}
&=&\left. \left\{ \frac{d^{2}\!\left\langle n\right\rangle }{dn_{\mathrm{g}%
}^{2}}\left( \frac{dn_{\mathrm{g}}}{du}\right) ^{2}+\frac{d\!\left\langle
n\right\rangle }{dn_{\mathrm{g}}}\frac{d^{2}n_{\mathrm{g}}}{du^{2}}\right\}
\right\vert _{0} \\
&=&\frac{d^{2}\!\left\langle n\right\rangle }{dn_{\mathrm{g}}^{2}}\frac{n_{%
\mathrm{NR}}^{2}}{\xi ^{2}}+\frac{d\!\left\langle n\right\rangle }{dn_{%
\mathrm{g}}}\frac{n_{\mathrm{NR}}}{\lambda }.  \notag
\end{eqnarray}%
Now we can use these expansions in Eq.~(\ref{Fq}). In what follows we are
interested in terms linear in $u$, since displacement-independent terms (we
name them $F_{\mathrm{q}0}$) result only in a constant displacement of the
resonator and do not influence the NR frequency and the quality factor:%
\begin{equation}
F_{\mathrm{q}}\approx F_{\mathrm{q}0}+u\left\{ \frac{V_{\mathrm{NR}}^{2}C_{%
\mathrm{NR}}}{2\lambda }-\frac{4E_{C}n_{\mathrm{NR}}^{2}}{\xi ^{2}}\alpha
\;+\left. \Xi \right\vert _{0}\right\} ,  \label{eq-tot}
\end{equation}%
where $\Xi $ is given by Eq. (\ref{Ksi2})

The first two terms in the brackets in Eq.~(\ref{eq-tot}) are of the form $%
\mathrm{const}\times u$. This results in constant shifts in the frequency
and quality factor, independent of the QD state. In contrast, the terms
denoted by $\Xi $ collect the QD-state dependent terms; these terms describe
the impact of the QD charge variations, $\delta \!\!\left\langle
n\right\rangle $, on the NR characteristics. In this way, when we obtained
Eq. (\ref{Ksi2}) in the main text, we meant \textquotedblleft keeping only
the terms defined by the QD state\textquotedblright , which assumed ignoring
the impact of the first two terms in the brackets in Eq.~(\ref{eq-tot}).

\end{document}